\documentclass[a4paper,11pt]{article}
\pdfoutput=1
\usepackage{jheppub}
\usepackage{pgfplots}
\usepackage{subcaption}
\pgfplotsset{compat=1.8}
\usepgfplotslibrary{groupplots}
\pgfplotsset{
    tick scale binop=\times
}
\pgfplotsset{every x tick label/.append style={font=\scriptsize, yshift=0.25ex},every y tick label/.append style={font=\scriptsize, xshift=0.25ex}, every major tick/.append style={major tick length=4pt, black}, every minor tick/.append style={minor tick length=2pt, black},every axis/.style={scale only axis},minor tick num = 4, xlabel near ticks,ylabel near ticks}

\usepackage{superfluid_preamble}

\title{Boost-invariant superfluid flows}

\author{Ronnie Rodgers and}
\author{Javier G. Subils}
\affiliation{
    Nordita, Stockholm University and KTH Royal Institute of Technology,\\
    Hannes Alfvéns väg 12, SE-106 91 Stockholm, Sweden
}
\emailAdd{ronnie.rodgers@su.se}
\emailAdd{javier.subils@su.se}

\preprint{NORDITA 2022-046}

\abstract{We present some exact solutions to the ideal hydrodynamics of a relativistic superfluid with an almost-conformal equation of state. The solutions have stress tensors which are invariant under Lorentz boosts in one direction, and represent superfluid generalisations of the Bjorken and Gubser flows. We also study corrections to the flows in first-order hydrodynamics, arguing that dissipation is dominated by the shear viscosity. We present some simple numerical solutions for these viscous corrections. Finally, we estimate the size of corrections to the flows arising when the spontaneously broken \(\mathrm{U}(1)\) symmetry responsible for superfluidity is only approximate, giving the corresponding Goldstone boson a small non-zero mass. We find that the massless solutions can still provide good approximations at sufficiently small spatial rapidities.}

\begin{document}

\maketitle

\section{Introduction}
\label{sec:intro}

Hydrodynamics is an effective field theory describing long wavelength, low frequency excitations of fluids, and is thus applicable to a diverse range of physical systems. In high energy physics, viscous hydrodynamics has proven to be an excellent framework for describing the experiments on heavy-ion collisions at both the Relativistic Heavy Ion Collider (RHIC) and the Large Hadron Collider (LHC)~\cite{Heinz:2013th,Jeon:2015dfa}. Furthermore, it has been claimed that the hydrodynamic approximation may even be helpful in collisions involving protons~\cite{Zhao:2017rgg,ALICE:2017smo}.

The hydrodynamic equations are difficult to solve, typically requiring numerical simulation. However, analytical solutions can sometimes be found for flows that preserve a large degree of symmetry. One such solution is the so-called ``Bjorken flow''~\cite{Bjorken:1982qr}. Bjorken flow arises when the colliding nuclei are modelled as infinite planes, leading to a flow that is invariant under Lorentz boosts along the beam axis, and translationally-invariant transverse to the beam axis. 

The latter approximation is of course unrealistic: the radius of a nucleus is on the order of 10\,fm, comparable to other scales of typical heavy-ion collisions (for instance, the hydrodynamisation time is of the order of $1\, \text{fm}/\text{c}$~\cite{Schlichting:2019abc}). A generalisation of Bjorken flow with no translational invariance in the transverse plane was found by Gubser in ref.~\cite{Gubser:2010ze}. However, this solution comes with its own unrealistic simplifications, such as that the initial state and dynamics respect relativistic conformal invariance, and that the collision is perfectly central.

Despite the highly simplifying assumptions of the Bjorken and Gubser flows, they have been extremely useful as a tool for building intuition and as a starting point for understanding more complicated regimes. In addition to providing analytical insight into the hydrodynamic equations, they also serve as a check for the numerical performance of hydrodynamic codes (for example, see ref.~\cite{Okamoto:2017ukz}). 

The aforementioned flows involve arise in the hydrodynamics of a normal fluid. However, if the up and down quarks were massless, quantum chromodynamics (QCD) would have a \(\mathrm{U}(1) \times \mathrm{SU}(2)_\mathrm{L} \times \mathrm{SU}(2)_\mathrm{R}\) global symmetry --- chiral symmetry --- which at low temperatures would be spontaneously broken to \(\mathrm{U}(1)\times\mathrm{SU}(2)_{\text{V}}\). In the presence of spontaneously broken continuous symmetries, new hydrodynamic modes appear (the Goldstone modes, which we will refer to as pions), and these should be included in the hydrodynamic equations~\cite{Son:2000ht}. The resulting fluid is called a superfluid.

In reality, the up and down quarks are massive, explicitly breaking chiral symmetry. However, since the scale of the explicit symmetry breaking is small, the corresponding Noether current is approximately conserved, and so a suitably modified form of superfluid hydrodynamics should still apply, with possible phenomenological implications. For example, refs.~\cite{Grossi:2020ezz,Grossi:2021gqi,Florio:2021jlx,Torres-Rincon:2022ssx} have advanced the critical dynamics near the pseudo-chiral phase transition as a possible explanation of an observed enhancement of soft pion production in Pb--Pb collisions~\cite{Mazeliauskas:2019ifr,Devetak:2019lsk,ALICE:2019hno,Nijs:2020roc,Guillen:2020nul}.

In this paper we will exploit the symmetry arguments used in refs.~\cite{Bjorken:1982qr,Gubser:2010ze} to find analytic solutions to the equations of ideal superfluid hydrodynamics. For simplicity we will consider only the spontaneous breaking of a \(\mathrm{U}(1)\) symmetry, and thus there will be only one pion. The solutions we find will also apply to systems with spontaneously broken non-abelian symmetries, with the \(\mathrm{U}(1)\) being the diagonal component of the full broken symmetry group. We leave generalisations of the solutions involving non-abelian effects to future work. We will also consider simple dissipative corrections to the analytic flows, which turn out to require some mild numerics.

Superfluid generalisations of Bjorken flow have been considered previously in the literature~\cite{Lallouet:2002th,Mitra:2020hbj}, and we should comment on the differences between our flows and those found in these earlier works. Ref.~\cite{Lallouet:2002th} found an analytic solution for superfluid Bjorken flow in low-temperature QCD by approximating the pion decay constant \(f\) by its zero temperature value \(f_\pi\). Accounting for a realistic temperature dependence of \(f\), ref.~\cite{Lallouet:2002th} also found some numerical solutions to the hydrodynamic equations. We extend these results by finding an analytic flow including the leading order low-temperature dependence of \(f\). Ref.~\cite{Mitra:2020hbj} included fluctuations of the chiral condensate (the sigma meson) in the list of hydrodynamic degrees of freedom, allowing their theory to apply also above and near the critical temperature. Unfortunately, the resulting equations are sufficiently complicated that only numerical solutions may be found.

In practice, the solutions we find are unlikely to be directly physically applicable to heavy-ion collisions. Both because we neglect fluctuations of the order parameter, and due to the simple equation of state we choose, the solutions will describe superfluid flows with local temperatures well below the approximate chiral symmetry breaking temperature~\(T_c\). In heavy-ion collisions, for \(T \lesssim T_c\) the hadrons chemically decouple~\cite{HEINZ1999140,Braun-Munzinger:2003htr}, and at yet lower temperatures they also kinetically decouple~\cite{Heinz:2004qz}, leading to a breakdown of the hydrodynamic description. However, we hope that the solutions we present are at least of theoretical interest, and that they may serve as a starting point for a more realistic treatment of the physics at temperatures closer to \(T_c\). Perhaps they may also be used to test future numerical hydrodynamic codes that account for superfluid contributions.

The rest of the paper is organised as follows. In section~\ref{sec:hydro} we will write down our model superfluid equation of state, and review the equations of superfluid hydrodynamics. In section~\ref{sec:bjorken} we will find both boost- and translationally-invariant solutions to these equations, mimicking Bjorken's treatment for a normal fluid. Flows with no translational invariance will be found in section~\ref{sec:gubser}, following Gubser's method. In section~\ref{sec:mass} we will estimate the size of corrections to our flows arising from small, non-zero pion mass. Finally, in section~\ref{sec:discussion} we will summarise our results and comment on possible directions for future work.

\section{Superfluid hydrodynamics}
\label{sec:hydro}

\subsection{Thermodynamics and ideal hydrodynamics}

In a normal fluid the hydrodynamic degrees of freedom are the densities of conserved charges, since they relax only slowly to equilibrium; a conserved charge relaxes over macroscopic timescales since it must spread throughout the system, while other quantities typically relax over microscopic timescales. In a superfluid, the gradient \(\p_\l \vf\) of the massless Goldstone boson also relaxes over macroscopic timescales, so must be included among the microscopic degrees of freedom, see e.g. ref.~\cite{Son:2000ht}.

In the superfluids of interest in this work, the conserved currents are the stress tensor \(T^{\m\n}\) and a \(\mathrm{U}(1)\) current \(J^\m\). Their equations of motion are the continuity equations
\begin{equation}
    \nabla_\m T^{\m\n} = \nabla_\m J^\m = 0.
    \label{eq:conservation_laws}
\end{equation}
In ideal superfluid hydrodynamics, the currents take the form~\cite{Son:2000ht}
\begin{align}
    T^{\m\n} &= \ve u^\m u^\n + p \D^{\m\n} + \m f^2 \le(u^\m \xi^\n + \xi^\m u^\n \ri) + f^2 \xi^\m \xi^\n,
    \nonumber \\
    J^\m &= \r u^\m + f^2 \xi^\m,
    \label{eq:ideal_hydro_tensors}
\end{align}
where \(u^\m\) is the normal fluid velocity, \(\D^{\m\n} \equiv g^{\m\n} + u^\m u^\n\) (with \(g_{\m\n}\) being an arbitrary metric) is a projector orthogonal to \(u^\m\), \(\x^\m \equiv \D^{\m\n} \p_\n\vf \) is proportional to the relative superfluid velocity, and \(\ve\), \(p\), \(\m\), \(\r\) and \(f^2\) are the energy density, pressure, chemical potential,  charge density, and pion susceptibility, respectively. In addition, we must impose the Josephson condition on the pion, which in ideal hydrodynamics reads
\begin{equation} \label{eq:ideal_josephson_condition}
    u^\l \p_\l \vf = -\m
    \qquad
    \Rightarrow
    \qquad
    \x^\l = \p^\l \vf - \m u^\l.
\end{equation}
The Josephson condition may be motivated in a number of different ways: it arises from the canonical conjugacy of the pion and the charge density~\cite{Son:2000ht}; it may be derived by demanding that the entropy current is conserved at the ideal order \cite{Herzog:2011ec,Jain:2016rlz}; finally, it arises as a natural consequence of a higher form symmetry associated to winding of the superfluid phase~\cite{Grozdanov:2018ewh,Delacretaz:2019brr}.

To complete the ideal hydrodynamic equations, we must specify an equation of state \(p(T,\m,\x)\), where \(\x \equiv \sqrt{\x^\l \x_\l}\) is real since \(\x^\l\) is spacelike (as it is by definition orthogonal to the timelike vector \(u^\m\)). The entropy density \(s\), charge density \(\r\), and pion susceptibility \(f^2\) are determined through the first law \(\diff p = s \, \diff T + \r \,\diff \m - f^2 \xi \, \diff \xi\), while the energy density is given by \(\ve = T s  + \m \r - p\). For clear discussions of the thermodynamics of superfluids, see refs.~\cite{Son:2000ht,PhysRevD.79.066002}.

Throughout this work, in order to obtain concrete solutions to the equations of superfluid hydrodynamics we will use a model equation of state of the form\footnote{Our equation of state is a special case of the more general equation of state for nuclear matter discussed in ref.~\cite{Son:1999pa}, obtained by treating the normal fluid component as being conformal.} 
\begin{equation} \label{eq:model_equation_of_state}
    p(T,\m,\xi) = n T^4 + \frac{1}{2} \chi(T) \m^2 - \frac{1}{2} f^2(T)\xi^2.
\end{equation}
The first term describes a normal fluid with conformal equation of state. This is one of the simplifying assumptions mentioned in the introduction that limits applications to QCD, in which the equation of state at \(\m = \x = 0\) takes a rather more complicated form~\cite{Philipsen:2012nu,HotQCD:2014kol}.\footnote{A conformal normal fluid may be more appropriate for models with chiral symmetry-broken phases that are deconfined, such as can occur in the holographic Sakai--Sugimoto model~\cite{Aharony:2006da}.}  The simplicity of the \(T^4\) term will be crucial in allowing us to derive exact solutions to the hydrodynamic equations.

The second and third terms in equation~\eqref{eq:model_equation_of_state} represent the leading corrections to the pressure at small non-vanishing chemical potential and superfluid velocity. The restriction to small chemical potential is merely a simplifying assumption, but it will at least turn out to be self-consistent; when we come to construct boost-invariant superfluid flows with this equation of state, we will find that they require us to set \(\m=0\). On the other hand, the restriction to small \(\x\) is physically motivated. At large \(\x\) superfluidity breaks down due to the creation of vortex filaments~\cite{Landau&Lifshitz,Landau&LifshitzSP2}. This has been observed in superfluid helium, for example~\cite{17c67fdc75ba4c0f9e13c247b78ccf97}.

We will take the susceptibilities in equation~\eqref{eq:model_equation_of_state} to have the form\footnote{Our motivation for the notation \(f_\pi\) is that in QCD, \(f(T=0) = \sqrt{\chi(T=0)}\) is the pion decay constant. In the QCD literature the susceptibility \(f\) is usually referred to as the spatial pion decay constant, and is often denoted \(f_s\), while \(\chi\) is the square of the temporal pion decay constant \(f_t\). See for example ref.~\cite{Son:2002ci}.}
\begin{equation} \label{eq:model_susceptibilities}
    f^2(T) = f_\pi^2 - k T^2,
    \qquad
    \chi(T) = f_\pi^2 - k' T^2.
\end{equation}
for some constants \((f_\pi,k,k')\). These are the leading order behaviours of the susceptibilities at low temperature in \(\mathrm{O}(N)\) linear sigma models, for example~\cite{Son:2002ci}. Since we are now also making a low temperature expansion, we expect the model equation of state~\eqref{eq:model_equation_of_state} with susceptibilities~\eqref{eq:model_susceptibilities} to be appropriate for superfluid flows with the hierarchy of scales
\begin{equation}
    \m, \x \ll T \ll f_\pi.
\end{equation}

The equation of state that we have chosen contains only one dimensionful parameter, \(f_\pi\), and so for \(f_\pi = 0\) the superfluid is conformal. Indeed, the energy density following from the equation~\eqref{eq:model_equation_of_state} is
\begin{equation} \label{eq:model_energy_density}
   \ve = 3 n T^4 + \frac{1}{2} \le(T \p_T \chi + \chi\ri) \m^2 - \frac{1}{2} \le(T \p_T f^2 - f^2 \ri) \xi^2,
\end{equation}
and we find the trace of the stress tensor to be \(T^\m{}_\m = -\ve + 3 p + f^2 \x^2 = - f_\pi^2 (\p \vf)^2\), which vanishes when \(f_\pi=0\). Notice that this expression for \(T^\m{}_\m\) is precisely what one would obtain for a free, massless scalar field \(\vf\) with action \( S = - f_\pi^2 \int \diff^4 x \, \sqrt{-g} \, (\p \vf)^2\). We will see that \(f_\pi^2\) plays very little role in the hydrodynamic equations, and so in section~\ref{sec:gubser} we will be able to leverage the powerful conformal-symmetry based method of ref.~\cite{Gubser:2012gy} to construct boost-invariant superfluid flows with non-trivial energy density profiles, even when \(f_\pi \neq 0\).

\subsection{First-order hydrodynamics}
\label{sec:first_order}

The first-order dissipative corrections to superfluid hydrodynamics are rather involved: the most general constitutive relations of first-order superfluid hydrodynamics contain twenty transport coefficients~\cite{Bhattacharya:2011tra}. We will consider only parity invariant superfluids, eliminating six of these coefficients~\cite{Bhattacharya:2011tra}. Further, the assumption of small relative superfluid velocity discussed above eliminates a further nine, leaving only the five transport coefficients discussed in ref.~\cite{Landau&Lifshitz}. Adopting the notation of ref.~\cite{Herzog:2011ec} for the dissipative corrections, one then obtains the constitutive relations\footnote{
    Our variables are related to the ones used in ref.~\cite{Herzog:2011ec} by  $ f^2 |_{\text{\tiny here}} = \frac{\rho_{\text{\tiny s}}}{\mus}|_{\text{\tiny there}} $ and $ \xi^\lambda  |_{\text{\tiny here}} = \mus n^\lambda|_{\text{\tiny there}}   $.
}
\begin{align}
    T^{\m\n} &= \Tideal^{\m\n} - \h \s^{\m\n} - \le(\z_1 \Delta^{\m\n}  + 2 \z_2 f^2 u^{(\m} \x^{\n)}\ri)\nabla_\l u^\l - \le(\z_2 \Delta^{\m\n}  + 2 \z_3 f^2 u^{(\m} \x^{\n)}\ri) \nabla_\l (f^2 \x^\l ),
    \nonumber \\[0.5em]
    J^\m &= \Jideal^\m - \k \Delta^{\m\n} \nabla_\n \le(\frac{\m}{T}\ri),
    \label{eq:viscous_hydro_tensors}
    \\[0.5em]
    u^\m \p_\m \vf &= - \m + \z_2 \nabla_\l u^\l + \z_3 \nabla_\l( f^2 \x^\l),
    \nonumber
\end{align}
where \(\Tideal^{\m\n}\) and \(\Jideal^\m\) are the ideal currents given in equation~\eqref{eq:ideal_hydro_tensors}, \(\{\h,\k,\z_{1,2,3}\}\) are the five transport coefficients, \(2 u^{(\m} \x^{\n)} \equiv u^\m \x^\n + \x^\m u^\n\), and
\begin{equation}
    \s^{\m\n} \equiv \D^{\m\a} \D^{\n\b} \le(\nabla_\a u_\b + \nabla_\b u_\a \ri) -  \frac{2}{3} \D^{\m\n}\nabla_\l u^\l.
\end{equation}

The five first-order transport coefficients are not completely free. Demanding that the entropy current has postive divergence imposes that $\eta, \k, \z_1, \z_3 \geq 0$ and \(\z_2^2 \leq \z_1 \z_3\)~\cite{Pujol:2002na,Gusakov:2007px,Herzog:2011ec}. Further, taking the trace of the stress tensor in equation~\eqref{eq:viscous_hydro_tensors}, one finds 
\begin{equation}
    T^\m{}_\m = (\Tideal)^\m{}_\m - 3 \z_1 \nabla_\l u^\l - 3 \z_2 \nabla_\l (f^2 \x^\l).
\end{equation}
We thus see that a conformally invariant theory has \(\z_1 = \z_2 = 0\). In section~\ref{sec:bjorken_dissipative} we will use this fact to argue that the shear viscosity \(\h\) is the most important first-order transport coefficient for the physics of superfluid Bjorken flow.

\section{Superfluid Bjorken flows}
\label{sec:bjorken}

\subsection{Ideal superfluid}

The natural coordinate system for studying Bjorken flow may be obtained starting from four-dimensional Minkowski space in cylindrical coordinates,
\begin{equation} \label{eq:metric_cylindrical}
    \diff s^2 = g_{\m\n} \diff x^\m \diff x^\n = - \diff t^2 + \diff z^2 + \diff \xp^2 + \xp^2 \diff \q^2,
\end{equation}
where we will refer to the \(z\) direction as the beam axis, while \((\xp,\q)\) are polar coordinates in the transverse plane. One then defines the Milne time \(\t\) and spatial rapidity \(w\) through \(t = \t \cosh w\) and \(z = \t \sinh w\). We use the symbol \(w\) for the spatial rapidity, rather than the more conventional \(\h\), in order to avoid confusion with the shear viscosity. The Milne time is invariant under boosts along the beam axis, while the rapidity is shifted to \(w \to w + \frac{1}{2} \log \le(\frac{1-v}{1+v}\ri)\), where \(v\) is the boost velocity. In the new coordinates, the metric~is
\begin{equation} \label{eq:metric}
    \diff s^2 = - \diff \t^2 + \t^2 \diff w^2 + \diff \xp^2 + \xp^2 \diff \q^2.
\end{equation}

We now seek solutions of the ideal superfluid hydrodynamic equations that are invariant under boosts in the \(z\) direction, rotations about the \(z\) axis, and translations perpendicular to \(z\). Together, these conditions imply that observables will depend only on \(\t\). Let us take the normal fluid velocity to be \(u = \p_\t\). The combination of the Josephson condition~\eqref{eq:ideal_josephson_condition} and symmetry then demand that \(\xi = \xi^w(\t) \p_w \). The fact that \(\p_\l \vf = \x_\l + \m u_\l\) is a gradient implies that \(\p_\t (\t^2 \xi^w) = 0\), so we may take \(\xi^w(\t) = \xh/\t^2\), for some constant \(\xh\). If we assume that \(\m = \m(\t)\), then we can straightforwardly solve \(\xi^\l = \p^\l \vf - \m u^\l\) for the Goldstone boson field, \(\vf = \xh w - \int \diff \t  \m(\t)\).

We can now evaluate the continuity equations~\eqref{eq:conservation_laws}. Conservation of the \(U(1)\) charge and the momentum, \(\nabla_\l J^\l = 0\) and \(\nabla_\l T^{\l w} = 0\) respectively, imply that
\begin{equation} \label{eq:bjorken_flow_charge_momentum_conservation}
    \p_\t (\t \mu \chi) = 0
    \qquad
    \text{and}
    \qquad
    \xh  \p_\t (\t \m f^2) = 0 .
\end{equation}
For non-zero \(\xh\) and \(\m\), this implies that \(\p_\t(\chi/f^2)\) = 0. Since  \(\chi\) and \(f^2\) are generically different functions of the local temperature~\cite{Son:2001ff,Son:2002ci}, we conclude that superfluid Bjorken flow is usually consistent only for \(\m = 0\). Concretely, for \(\chi\) and \(f^2\) given by equation~\eqref{eq:model_susceptibilities}, equations~\eqref{eq:bjorken_flow_charge_momentum_conservation} imply that \(\m =0\) unless \(k = k'\). We will therefore set the local chemical potential to zero for the remainder of this section, satisfying the charge and momentum conservation equations~\eqref{eq:bjorken_flow_charge_momentum_conservation} trivially. The solution for the Goldstone mode is then~\(\vf = \xh w\).

A notable example of a theory with \(k = k'\) is actually chiral perturbation theory, where it has been shown that \(f_\pi^2\) and \(\chi\) differ by a term \(\cO(T^4)\) at low temperatures~\cite{Pisarski:1996mt,Toublan:1997rr}. Thus, at the level of the low-temperature approximation used in equation~\eqref{eq:model_susceptibilities}, the equation of state from chiral perturbation theory permits superfluid Bjorken flow with non-zero \(\m\). Unfortunately, we have not been able to find an analytic solution for the \(\m \neq 0\) case.\footnote{Note that \(\m=0\) still provides a consistent solution to equation~\eqref{eq:bjorken_flow_charge_momentum_conservation} even when \(k=k'\), so our results are still applicable to this case.} This case is studied numerically in ref.~\cite{Lallouet:2002th}.

With \(\m=0\), the only non-trivial conservation law is the conservation of energy, \(\nabla_\l T^{\l \t} = 0\), which implies
\begin{equation}
    \frac{1}{\t^{1/3}} \p_\t \le(\t^{1/3} T\ri) + \frac{k \xh^2 \p_\t(T/\t)}{12 n \t T^2} = 0 .
\end{equation}
The general solution is
\begin{equation} \label{eq:ideal_bjorken_temperature}
	T(\t) = \frac{3 n \S(\t)^{2/3} -  k \xh^2}{6 n \t \S(\t)^{1/3}},
	\qquad
	\S(\t) = \frac{4 \t^2}{\t^2_0}  + \sqrt{\frac{16 \t^4}{\t^4_0}  + \frac{ k^3 \xh^6}{27n^3}},
\end{equation}
where \(\t_0\) is an integration constant. Substituting this solution into equation~\eqref{eq:model_energy_density} we obtain the corresponding energy density,
\begin{equation} \label{eq:ideal_bjorken_energy_density}
    \ve(\t) = 3 n \le[\frac{3n \S(\t)^{2/3} -   k \xh^2}{6 n \t \S(\t)^{1/3}}\ri]^4 + \frac{k \xh^2}{2 \t^2} \le[\frac{3 n \S(\t)^{2/3} -   k \xh^2}{6 n \t \S(\t)^{1/3}}\ri]^2 + \frac{f_\pi^2 \xh^2}{2 \t^2}.
\end{equation}
At late times, this energy density decays as
\begin{equation} \label{eq:ideal_bjorken_energy_density_late_times}
	\ve(\t)  = \frac{3 n}{ {\t_0}^{8/3}  \t^{4/3}} + \frac{f_\pi^2 \xh^2}{2 \t^2} - \frac{k \xh^2}{2 {\t_0}^ {4/3}  \t^{8/3}} + \dots \; .
\end{equation}
The first term in the expansion is the usual Bjorken flow energy density, with superfluidity providing subleading corrections. Notice that the zero temperature pion decay constant \(f_\pi\) appears only in the first subleading correction to the late-time energy density, which decays as \(\t^{-2}\). We then obtain a series of further corrections decaying as larger powers of \(\t^{-1}\), controlled by the finite temperature correction to the susceptibility \(f^2\).

The term in equation~\eqref{eq:ideal_bjorken_energy_density_late_times} proportional to \(\t^{-2}\) is reminiscent of a similar term found in the superfluid Bjorken flow of ref.~\cite{Lallouet:2002th}, however it has a different origin. Their flow has \(\xh = 0\), but a non-zero chemical potential \(\m \propto \t^{-1}\) (allowed because they set \(f^2 =  \chi = f_\pi^2\)). This leads to an energy density containing a term \(\propto \m^2 \propto \t^{-2}\).

Expanding equation~\eqref{eq:ideal_bjorken_temperature} at small proper time \(\t \to 0\), one finds that the local temperature grows linearly as \(T(\t) \approx 4 n \t /( k \xh^2 \t_0^2)\) for non-zero \(\xh\). Thus, the system appears to heat up with \(\t\) at early times, and then cool down at late times. However, this may just be an artifact of our definition of the local temperature. In particular, no such heating is apparent in the energy density, which also decreases at early times as
\begin{equation}
	\ve(\t) \approx \frac{f_\pi^2 \xh^2}{2 \t^2} + \frac{8 n^2}{k \xh^2 \t_0^2} + \dots \; .
\end{equation}

Finally, we make a brief comment on the behaviour of the \(\mathrm{U}(1)\) current \(J^\m\) in our solutions. In the fluid frame, the charge density \(u_\m J^\m = 0\), while there is a non-zero current in the \(w\)-direction that decays with increasing \(\t\). The situation is rather more intricate in the ``laboratory frame''~\eqref{eq:metric_cylindrical}. In cylindrical coordinates \((t,z)\) the current is
\begin{equation} \label{eq:ideal_bjorken_current}
	J = \le(f_\pi^2  - k T^2 \ri) \le(\frac{z \, \p_t + t \, \p_z}{t^2 - z^2} \ri),
\end{equation}
with \(T\) given by equation~\eqref{eq:ideal_bjorken_temperature} (and \(\t^2 = t^2 - z^2\)). At late Minkowski time \(t\) the contribution from \(T^2\) is negligible, and the current~\eqref{eq:ideal_bjorken_current} takes a rather simple form \(J = f_\pi^2 (z \, \p_t + t \, \p_z)/(t^2-z^2)\). For fixed \(t\), we see that the current density diverges as \(J^t \to \pm \infty\) as the lightcone at \(z = \pm t\) is approached. The solution thus describes two shockwaves of equal and opposite infinite charge density travelling outwards at the speed of light. The total charge in the \(z \geq 0\) region increases with \(t\), due to the non-zero \(J^z\). At early and intermediate \(t\), the behaviour of the current is rather more intricate due to the temperature dependence in equation~\eqref{eq:ideal_bjorken_current}. In particular, the charges of the shockwaves can flip sign at intermediate times.

\subsection{Dissipative corrections}
\label{sec:bjorken_dissipative}

We would now like to study dissipative corrections to the flow in equation~\eqref{eq:ideal_bjorken_energy_density}. As for the ideal case, the assumed symmetries force observables to be functions only of \(\t\). We continue to take the normal fluid velocity to be \(u = \p_\t\), and consequently the relative superfluid velocity is still of the form \(\xi = \xi^w(\t) \p_w \). With this ansatz, two of the five transport coefficients discussed in section~\ref{sec:first_order} have no effect on the flow: it is straightforward to show that when \(T\) and \(\m\) are functions only of \(\t\),
\begin{equation} \label{eq:vanishing_transport_terms}
	\nabla_\m \le[f(T)^2 \x^\m\ri] = \Delta^{\m\n} \nabla_\n \le(\frac{\m}{T}\ri) = 0.
\end{equation}
Comparing to the first-order constitutive relations in equation~\eqref{eq:viscous_hydro_tensors}, we see that \(\z_3\) and \(\k\) drop out of the hydrodynamic equations completely.

We are then left with three transport coefficients: the shear viscosity \(\h\), the bulk viscosity \(\z_1\), and \(\z_2\). In principle, we would need to do a microscopic calculation to compute how these transport coefficients depend on \(T\), \(\m\) and \(\x\). However, as noted in section~\ref{sec:first_order}, for a conformal superfluid \(\z_1 = \z_2 = 0\). Although our superfluid is not conformal, the only dimensionful parameter is \(f_\pi\), which drops out of all equations when \(\m = \x = 0\). We therefore expect \(\z_{1,2} = 0\) in this limit.\footnote{More precisely, \(\z_2\) may not be zero, but the combination \(\z_2 f^2 u^{(\m} \x^{\n)} \) appearing in the stress tensor in equation~\eqref{eq:viscous_hydro_tensors} vanishes in the \(\x=0\) limit. A non-zero, \(\t\)-dependent \(\z_2\) will provide a contribution to \(u^\m \p_\m \vf\) even at \(\x=0\) via the Josephson condition on the last line of equation~\eqref{eq:viscous_hydro_tensors}, but this will not affect the conservation equations of \(T^{\m\n}\) and \(J^\m\), which depend on \(\vf\) only through the transverse gradient \(\Delta^{\m\n} \p_\n \vf\).} In the case of interest, namely \(\m = 0\) and small non-zero \(\x\), we thus expect the shear viscosity \(\h\) to be much larger than \(\z_{1,2}\). For the remainder of this section, we will therefore set \(\z_{1,2}= 0\) and concentrate on the effects of the shear viscosity. We will further approximate the shear viscosity by its conformal form \(\h \approx \bar{\h} T^3\) where \(\bar{\h}\) is a dimensionless constant, neglecting any corrections at non-zero \(\x\).

\begin{figure}
	\definecolor{plotpurple}{rgb}{0.5,0,0.5}
	\begin{subfigure}{\textwidth}
	\begin{center}
		\hspace{25pt}\includegraphics{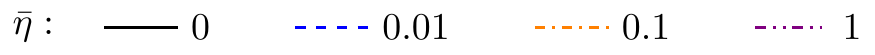}
		\vspace{-0.5cm}
		\includegraphics{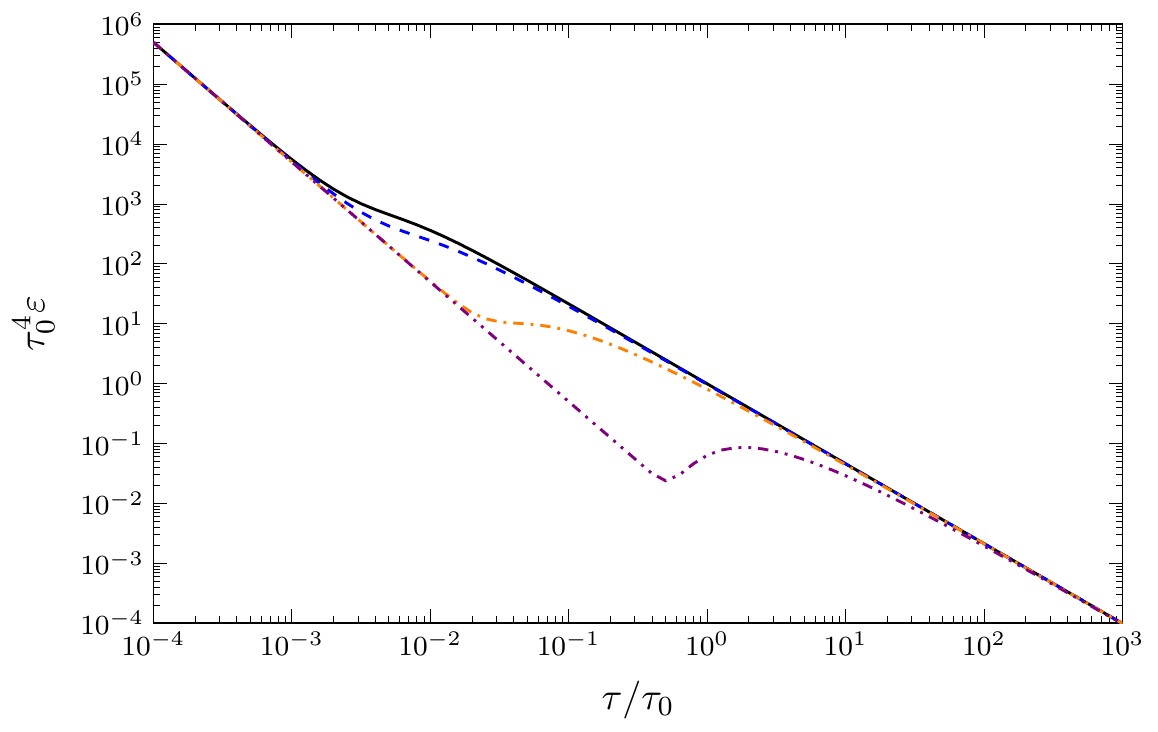}
	\end{center}
	\caption{Energy density}
	\label{fig:viscous_temperature_notsubtracted}
	\end{subfigure}
	\begin{subfigure}{\textwidth}
	\vspace{0.5cm}
	\begin{center}
		\includegraphics{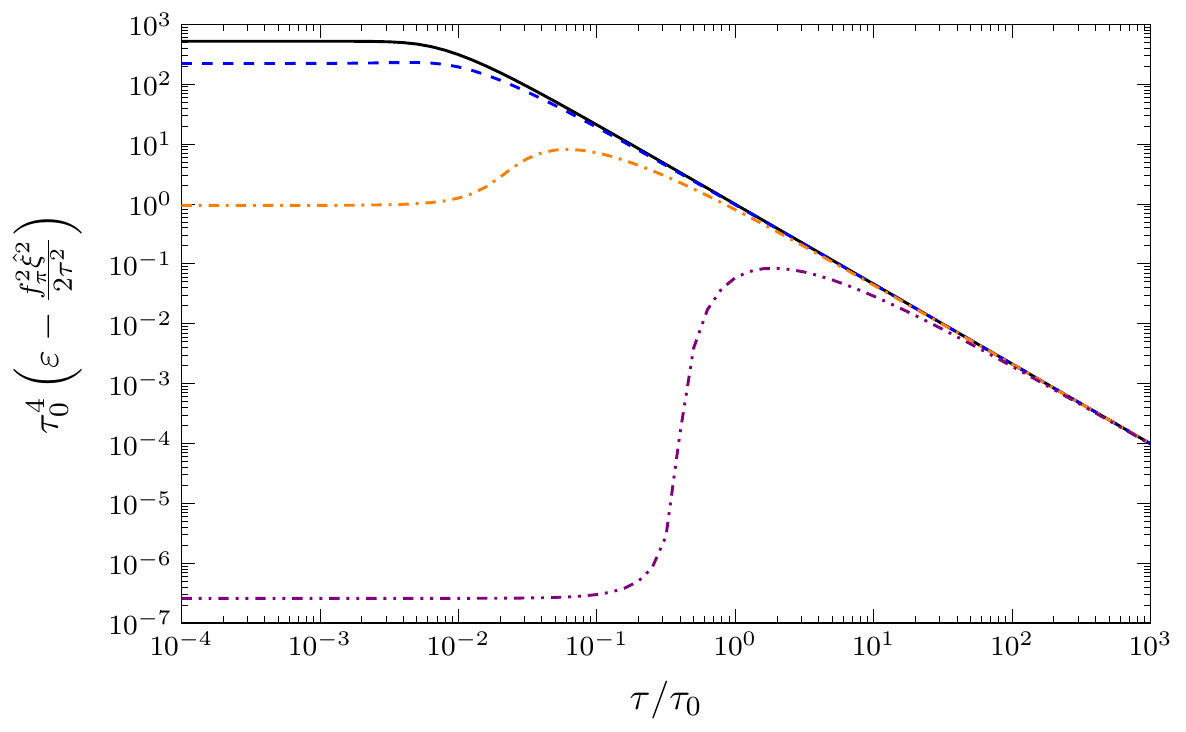}
	\caption{Subtracted energy density}
	\label{fig:viscous_temperature_subtracted}
	\end{center}
	\end{subfigure}
		\caption{\textbf{(a)} Numerical results for the energy density as a function of $\tau$ for superfluid Bjorken flow with non-zero shear viscosity \(\h\). The different curves correspond to different values of \(\bar{\h} \equiv \h/T^3\). To make the plot, we have chosen the parameters in the equation of state and the integration constants to be \(n = \pi^2/30\), \(k = 1/6\), \(\xh = 1/10\), and \(\t_0 = 1/f_\pi\). \textbf{(b)} The same data as in (a), but with the dominant early-time behaviour \(\ve \approx f_\pi^2 \xh^2/2\t^2\) subtracted, highlighting the difference between the curves for different \(\bar{\h}\).}
		\label{fig:viscous_temperature}
\end{figure}

The net effect of these approximations is that we are keeping terms up to \(\cO(\x^2)\) in ideal hydrodynamics, while throwing away everything except the normal fluid contribution to first-order hydrodynamics. In principle, given formulas for \(\h\), \(\z_1\), and \(\z_2\) at non-zero \(\x\), there is no barrier to extending the treatment in this section to account for dissipative superfluid corrections to the flow. However, we expect these corrections to be doubly small in the physically interesting regime, being suppressed in both the small-\(\x\) and derivative expansions.
With \(\m=0\) and the only non-trivial transport coefficient being \(\h = \bar{\h}T^3\), the Josephson condition implies \(\vf = \xh w\) for some dimensionless constant \(\xh\), as in the ideal case, while the only non-trivial conservation equation is the conservation of energy \(\nabla_\lambda T^{\lambda\tau} = 0\), which implies
\begin{equation} \label{eq:viscous_conservation_energy}
	\frac{1}{\t^{1/3}} \p_\t \le(\t^{1/3} T\ri) + \frac{k \xh^2 \p_\t(T/\t)}{12 n \t T^2} - \frac{\bar{\eta} T^2}{9n \tau^2}= 0.
\end{equation}
This equation is solved implicitly by the solution to the algebraic equation
\begin{equation}\label{eq:viscous_solution_T}
	\frac{4 \bar{\h}}{\sqrt{36 n k \xh^2 - \bar{\h}^2 }}\arctan\left( \frac{\sqrt{36 n \xh^2 - \bar{\h}^2 }}{\bar{\h} + 12 n T(\tau) \tau}\right)   +  \log \left[\t_0^2T(\tau)^3\left(\frac{\bar{\h}}{6 nT(\tau)}  +  \frac{k \xh^2 }{4 n  \tau T(\tau)^2 }  +  \tau \right)\right] = 0  ,
\end{equation}
where \(\t_0\) is an integration constant, chosen such that the leading order behaviour of \(T(\t)\) at late times matches that of the inviscid solution~\eqref{eq:ideal_bjorken_temperature}. We cannot solve this transcendental equation~\eqref{eq:viscous_solution_T} for $T(\tau)$ analytically, but it is straightforward to obtain numerical solutions, for example with the Newton--Raphson method. Substituting the resulting temperature into equation~\eqref{eq:model_energy_density}, we then obtain numerical results for the energy density as a function of \(\t\). For example, in figure~\ref{fig:viscous_temperature} we show results for the energy density for various different values of \(\bar{\h}\), for the following sample values of the parameters: \(n = \pi^2/30\), \(k = 1/6\), \(\xh = 1/10\), and \(\t_0 = 1/f_\pi\).\footnote{The choices of \(n\) and \(k\) were motivated by \(\mathrm{O}(N)\) sigma models, for which \(n = (N-1) \pi^2/90\) and \(k = (N-2)/12\)~\cite{kapusta_gale_2006}. We obtain the chosen values of \(n\) and \(k\) by taking \(N=4\), as relevant for QCD. The value \(k=1/6\), which is also found in chiral perturbation theory~\cite{Pisarski:1996mt,Toublan:1997rr}. The qualitative behaviour of \(\ve(\t)\) does not depend sensitively on the choices of \(n\) and \(k\).}

At small \(\t\) the curves appear to coincide in figure~\ref{fig:viscous_temperature_notsubtracted} at \(\ve \approx f_\pi^2 \xh^2/2\t^2\), due to the dominance of the final term in the energy density~\eqref{eq:ideal_bjorken_energy_density}. At large \(\t\) the curves coincide again, as can be seen from the following argument. We can solve equation~\eqref{eq:viscous_conservation_energy} perturbatively in \(1/\t\) to determine the late-time behaviour of the local temperature. Substituting the result into equation~\eqref{eq:model_energy_density}, we find that the energy density decays at late times as
\begin{equation} \label{eq:bjorken_viscous_late_time}
	\varepsilon(\tau)  = \frac{3n}{\t_0^{8/3}\tau^{4/3}} +  \frac{\t_0^2f_\pi^ 2 \xh^2 -4 \bar{\h} }{2\t_0^2\tau^ 2} -  \frac{n k \xh^ 2 - \bar{\h}^2}{2n \t_0^{4/3} \tau^ {8/3}}+ \dots \;.
\end{equation}
We see that the leading order correction to \(\ve(\t)\) due to non-zero shear viscosity occurs at the same order in the late-time expansion as the leading-order correction from non-zero \(\x\).

Finally, we note that the superfluid component regularises a singularity present in the Bjorken flow of a viscous normal fluid. The normal fluid local temperature, obtained by solving equation~\eqref{eq:viscous_conservation_energy} with \(\xh=0\), is
\begin{equation}
	T(\t) = \frac{1}{\t_0^{2/3}\t^{1/3}} - \frac{\bar{\h}}{6 n \t}.
\end{equation}
As pointed out in ref.~\cite{Gubser:2010ze}, for example, this local temperature becomes negative at early times, when the second term dominates over the first. This is not a huge problem, it is just a symptom of the expected breakdown of hydrodynamics at early times when gradients become large. However, it is perhaps pleasing that no such singularity exists in the superfluid case, as visible from the lack of a zero in the subtracted energy densities plotted in figure~\ref{fig:viscous_temperature_subtracted}. Concretely, solving equation~\eqref{eq:viscous_conservation_energy} perturbatively at small \(\t\), we find that at leading order the local temperature scales linearly in \(\t\) as

\begin{equation}
	T(\t) \approx \frac{4 n}{k \xh^2 \t_0^2} \le(\frac{\bar{\h} - \sqrt{\bar{\h}^2 - 36 n k \xh^2}}{\bar{\h} + \sqrt{\bar{\h}^2 - 36 n k \xh^2}} \ri)^{2 \bar{\h}/\sqrt{\bar{\h}^2- 36 n k \xh^2}} \t + \dots \; .
\end{equation}
The proportionality coefficient is (perhaps contrary to first appearance) real and positive for all real \(\bar{\h}\).

However, notice that at intermediate \(\t\) there is a very large deviation between the \(\bar{\h}=0\) curve in figure~\ref{fig:viscous_temperature} and those corresponding to all but the smallest \(\bar{\h} \neq 0\). The zero- and first-order terms in the hydrodynamic expansion are thus comparable in this regime, so we cannot trust hydrodynamics along the whole flow except for small values of \(\bar{\h}\).

\section{Transversely expanding flows}
\label{sec:gubser}

\subsection{Review of the normal fluid case}
\label{sec:normal_gubser}

In the previous section we described a boost-invariant superfluid flow that is also invariant under translations transverse to the beam axis. For real applications, for example to heavy-ion collisions, this translational invariance is unrealistic, since the fluid will be expanding from some central collision region.

In ref.~\cite{Gubser:2010ze}, Gubser describes an elegant method for deriving simple, transversely expanding, boost-invariant flows of normal, uncharged fluids with conformal equation of state \(p = \ve/3\). Gubser's method works by replacing the transverse translational invariance with a generalised conformal invariance. In this subsection we will review Gubser's method. In section~\ref{sec:superfluid_gubser} we will show how the method may be applied to superfluids.

The first step is to define a vector field~\cite{Gubser:2010ze}
\begin{align}
    \z &= a + q^2 b
    \nonumber \\
    &= 2 q^2 \t \xp \cos\q \, \p_\t + (1 + q^2 \t^2 + q^2 \xp^2) \cos\q \, \p_{\xp} - \frac{1 + q^2 \t^2 - q^2 \xp^2}{\xp} \sin \q \, \p_\q,
    \label{eq:conformal_killing_vector}
\end{align}
where \(q\) is a constant with dimensions of inverse length, \(a\) is the vector generating a unit translation in the \(\q = 0\) direction, and \(b\) is the vector generating a unit special conformal transformation in the same direction. Thus, \(\z\) is a conformal Killing vector of the Minkowski metric~\eqref{eq:metric},
\begin{equation} \label{eq:conformal_killing_vector_eq}
    \cL_\z g_{\m\n} = \frac{1}{2} \le(\nabla_\l \z^\l \ri) g_{\m\n},
\end{equation}
where \(\cL_\z\) is the Lie derivative along \(\z\).

Given a tensor field \(\mathcal{T}^{\n_1 \n_2 \dots}_{\m_1 \m_2 \dots}\), we say that it transforms covariantly under \(\z\) if it satisfies
\begin{equation} \label{eq:zeta_transformation_law}
    \cL_\z \mathcal{T}^{\n_1 \n_2 \dots}_{\m_1 \m_2 \dots} = - \frac{\a[\mathcal{T}^{\n_1 \n_2 \dots}_{\m_1 \m_2 \dots} ]}{4} \le(\nabla_\l \z^\l \ri) \mathcal{T}^{\n_1 \n_2 \dots}_{\m_1 \m_2 \dots} \; ,
\end{equation}
where the \textit{weight} \(\a[\mathcal{T}^{\n_1 \n_2 \dots}_{\m_1 \m_2 \dots} ]\) is a real constant. For instance, equation~\eqref{eq:conformal_killing_vector_eq} implies that the metric has weight \(\a[g_{\m\n}] = -2\). The inverse metric \(g^{\m\n}\) correspondingly has \(\a[g^{\m\n}]=2\), ensuring that the Kronecker delta \(\d^\m_\n = g^{\m\l} g_{\n\l}\) has vanishing weight.

Gubser's flows are constructed by demanding that the fluid velocity and energy density transform covariantly under \(\z\). Applying \(\cL_\z\) to the normalization condition \(g^{\m\n}u_\m u_\n = -1\) and using \(\a[g^{\m\n}] =2\), it is straightforward to show that the fluid velocity must have weight \(\a[u_\m] = -1\), so that
\begin{equation} \label{eq:fluid_velocity_zeta_transformation_law}
    \cL_\z u_\m = \frac{1}{4}  \le(\nabla_\l \z^\l \ri) u_\m.
\end{equation}
Assuming rotational invariance around the beam axis --- implying that \(u_\q = 0\) --- and boosting to a frame where \(u_w = 0\), the solution to the differential equation~\eqref{eq:fluid_velocity_zeta_transformation_law} is~\cite{Gubser:2010ze}
\begin{equation} \label{eq:gubser_normal_fluid_fluid_velocity}
    u = \cosh \k \, \p_\t + \sinh \k \, \p_{\xp},
    \qquad
    \k = \arctanh \le(\frac{2 q^2 \t \xp}{1 + q^2 \t^2 + q^2 \xp^2}\ri).
\end{equation}

Having fixed the fluid velocity, we now turn to the conservation of the stress tensor. If we demand that the energy density transforms covariantly under \(\z\) with weight \(\a[\ve] \equiv \a_\ve\), this implies~\cite{Gubser:2010ze}
\begin{equation} \label{eq:gubser_normal_fluid_energy_density_form}
    \ve = \frac{\hat{\ve}(g)}{\t^{\a_\ve}},
    \qquad
    g = \frac{1 - q^2 \t^2 + q^2 \xp^2}{2q\t},
\end{equation}
where \(\hat{\ve}(g)\) is a so far arbitrary function. The variable \(g\) is the only independent scalar combination of \(\t\) and \(\xp\) satisfying \(\cL_\z g =0\), up to trivial redefinitions, so any scalar with vanishing weight must be a function of \(g\) only. Assuming a conformal equation of state \(p = \ve/3\), at the level of ideal hydrodynamics the stress tensor is \(T^{\m\n} = \ve \le(4 u^\m u^\n + g^{\m\n} \ri)/3\). With the fluid velocity given by equation~\eqref{eq:gubser_normal_fluid_fluid_velocity} and the energy density taking the form~\eqref{eq:gubser_normal_fluid_energy_density_form}, conservation of the stress tensor implies that the energy density satisfies
\begin{equation} \label{eq:gubser_normal_fluid_energy_density_equation}
    3(1+g^2)\hat{\ve}'(g) + 4 g (\a_\ve-2) \hat{\ve}(g) + 4 q \t (\a_\ve - 4) \hat{\ve}(g)  = 0.
\end{equation}
Plainly this equation is only consistent with the assumption that \(\hat{\ve}(g)\) is a function only of \(g\) and not \(\t\) if \(\a_\ve = 4\). This value may also be obtained from dimensional analysis: if \(\hat{\ve}(g)\) is dimensionless, then \(\a_\ve = 4\) correctly assigns the energy density dimensions of \((\mathrm{length})^{-4}\). We may straightforwardly solve equation~\eqref{eq:gubser_normal_fluid_energy_density_equation} with \(\a_\ve=4\), and substitute into equation~\eqref{eq:gubser_normal_fluid_energy_density_form} to find the energy density~\cite{Gubser:2010ze}
\begin{equation} \label{eq:gubser_normal_fluid_energy_density_solution}
    \ve(\t,\xp) = \frac{\hat{\ve}_0}{(2q)^{8/3}\t^4 (1 + g^2)^{4/3}} = \frac{\hat{\ve}_0}{\t^{4/3}\le[1 + 2 q^2(\t^2 + \xp^2) + q^4 (\t^2 - \xp^2)^2\ri]^{4/3}},
\end{equation}
where \(\hat{\ve}_0\) is an integration constant.

Ref.~\cite{Gubser:2010ze} also computed the first-order dissipative correction to equation~\eqref{eq:gubser_normal_fluid_energy_density_solution}. Since the fluid is assumed to have conformal equation of state, the bulk viscosity vanishes and so the only non-zero first-order transport coefficient is the shear viscosity \(\h\), which contributes to the stress tensor as in equation~\eqref{eq:viscous_hydro_tensors}. Further, by dimensional analysis, the shear viscosity in the conformal fluid must take the form \(\h = \bar{\h}\ve^{3/4} \) for some dimensionless coefficient \(\bar{\h}\). The stress tensor therefore reads
\begin{equation}
    T^{\m\n} = \frac{\ve}{3} \le(4 u^\m u^\n + g^{\m\n} \ri)- \bar{\h} \ve^{3/4}  \, \sigma^{\mu\nu},
\end{equation}
with the fluid velocity given by equation~\eqref{eq:gubser_normal_fluid_fluid_velocity}. Conservation of this stress tensor leads to a modified version of equation~\eqref{eq:gubser_normal_fluid_energy_density_equation}, with solution~\cite{Gubser:2010ze}
\begin{equation}\label{eq:energy_Gubser_viscous}
    \ve(\t,\xp) = \frac{1}{\t^4} \le\{ \frac{\ve_0^{1/4}}{(1+ g^2)^{1/3}} + \frac{\bar{\h} g}{ \sqrt{1+g^2}} \le[1 - (1+g^2)^{1/6} {}_2 F_1\le(\frac{1}{2},\frac{1}{6}; \frac{3}{2}; -g^2\ri) \ri]\ri\}^4,
\end{equation}
where we leave implicit that \(g\) is the function of \(\t\) and \(\xp\) appearing in equation~\eqref{eq:gubser_normal_fluid_energy_density_form}.

\subsection{Superfluid case, ideal hydrodynamics}
\label{sec:superfluid_gubser}

We will now show how to generalise the conformal symmetry based-approach reviewed in section~\ref{sec:normal_gubser} to a superfluid with equation of state~\eqref{eq:model_equation_of_state}. We will once again seek flows in which all physical quantities transform covariantly under \(\z\) as in equation~\eqref{eq:zeta_transformation_law}. The procedure for determining the normal fluid velocity is unchanged, so that \(u\) is still given by equation~\eqref{eq:gubser_normal_fluid_fluid_velocity}. We then fix the superfluid velocity \(\x\) by demanding that it transforms as in equation~\eqref{eq:zeta_transformation_law} with weight \(\a[\x^\m] \equiv \a_\x\). Combining this with the condition that \(u_\m \x^\m = 0\) that arises from the definition \(\x^\m \equiv \Delta^{\m\n} \p_\n \vf\), one finds
\begin{equation} \label{eq:superfluid_velocity_general}
    \x = \frac{\hat{\x}(g)}{\t^{\a_\x}} \p_w.
\end{equation}
We now impose the condition that \(\p_\l \vf = \x_\l + \m u_\l\) is a gradient, which implies that \(\p_\m(\x_\n + \m u_\n) - \p_\n (\x_\m + \m u_\m) =0\). If we assume that \(\m\) transforms with weight \(\a[\m] \equiv \a_\m\), so that \(\m = \mh(g)/\t^{\a_\m}\), then the various components of this gradient condition imply
\begin{equation}
    (\a_\x-2)\xh(g) = 0,
    \qquad
    \le(\a_\m - 1 \ri)\mh(g)  = 0,
    \qquad
    \xh'(g)  = 0.
\end{equation}
Thus, non-zero solutions for \(\xh\) and \(\mh\) are only possible if \(\a_\x = 2\) and \(\a_\m = 1\), respectively, and the only possible solution for \(\xh\) is that it is constant.

Now we can attempt to solve the conservation equations. We will assume that the local temperature transforms covariantly under \(\z\) with weight \(\a_T\), so that we may write \(T = \Th(g)/\t^{\a_T}\). As for the Bjorken flow described in section~\ref{sec:bjorken}, we find that simultaneous conservation of \(J^\l\) and \(T^{\l w}\) is impossible unless the local chemical potential vanishes. Concretely, taking a linear combination of the equations \(\nabla_\l J^\l = 0\) and \(\nabla_\l T^{\l w} = 0\), we find the condition
\begin{equation}
    (k-k') \xh \, \Th(g) \mh(g) \le[ (1+g^2 ) \Th'(g) + \a_T (g + q \t) \Th(g)\ri] = 0.
\end{equation}
This equation makes clear that for the generic case \(k \neq k'\), flows with non-trivial superfluid component \(\xh \neq 0\) and non-trivial local temperature are only possible if \(\mh(g) = 0\). We then find that the conservation equations for the remaining components of the stress tensor are only consistent with the assumption that \(\Th(g)\) is a function only of \(g\) if \(\a_T = 1\), which is also what one would expect from dimensional analysis. In this case, conservation of the stress tensor implies
\begin{equation} \label{eq:superfluid_bjorken_energy_conservation}
    (1+g^2) \le[12 n \Th(g)^2 + k \xh^2\ri] \Th'(g) + 2 g \le[4 n \Th(g)^2 + k \xh^2 \ri]\Th(g) = 0.
\end{equation}

\begin{figure}
    \begin{center}
    \hspace{45pt}\includegraphics{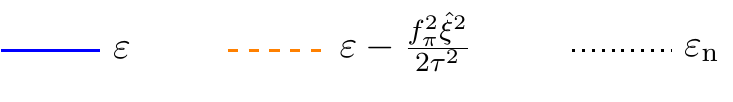}
    \vspace{-0.5cm}
    \includegraphics{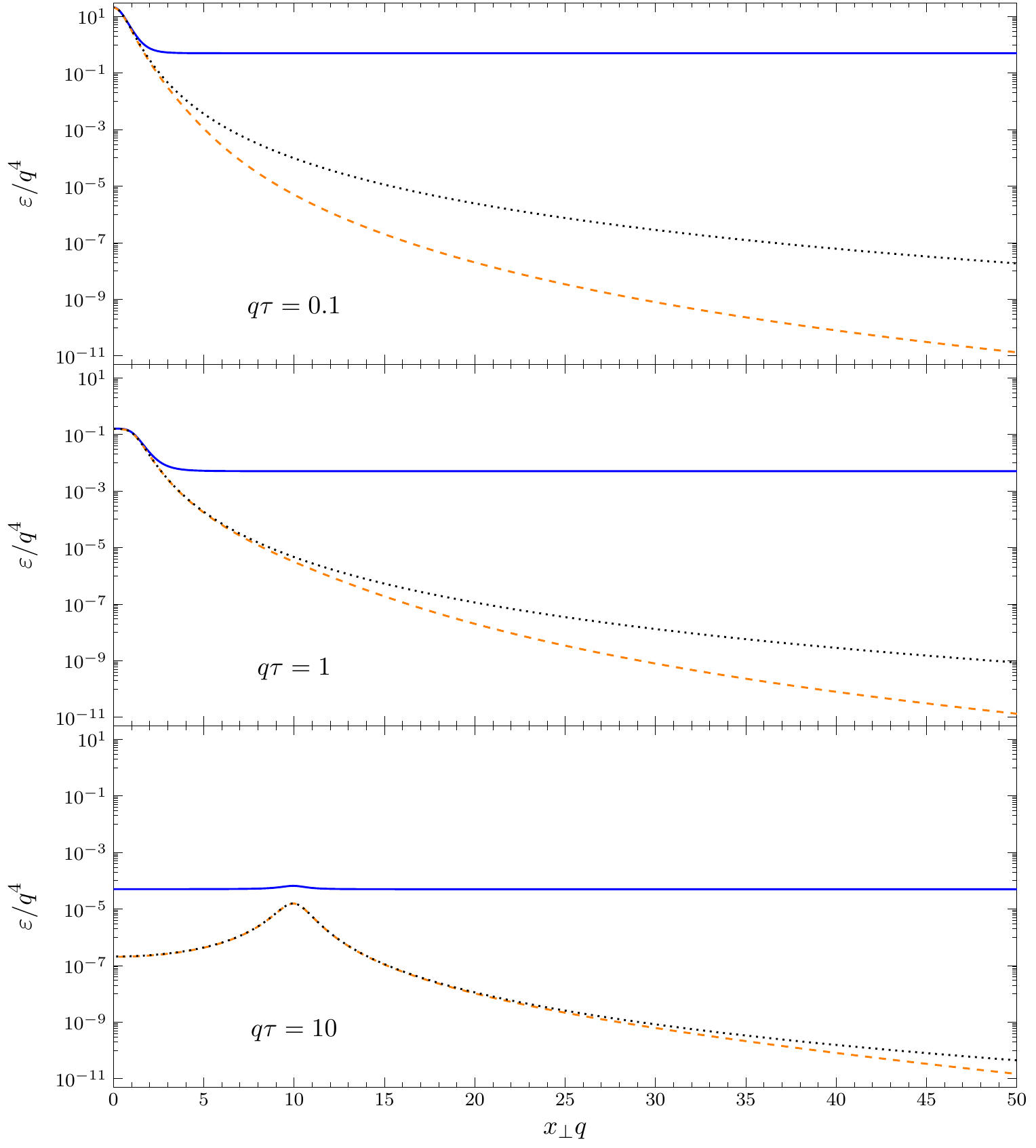}
    \end{center}
    \caption{Sample logarithmic plots of the energy density of ideal superfluid Gubser flow as a function of transverse distance \(\xp\), at different Milne times \(\t\). To make these plots, we have fixed the parameters of the model and the solution to \(n = \pi^2/30\), \(k = 1/6\), \(q = f_\pi\), \(\xh = 1/10\), and \(\t_0 = 1/f_\pi\). The solid blue curves show the energy density~\eqref{eq:gubser_superfluid_energy_density_solution}. The dashed orange curves show the same energy density~\eqref{eq:gubser_superfluid_energy_density_solution}, but with the \(\xp\)-independent contribution \(f_\pi^2 \xh^2/2\t^2\) subtracted, in order to make the \(\xp\)-dependence clearer. The dotted grey curves show the normal fluid result~\eqref{eq:gubser_normal_fluid_energy_density_solution}, with \(\hat{\ve}_0\) chosen such that the two results agree at \(q\t=0.1\) and \(\xp = 0\), for comparison.}
    \label{fig:example_gubser_energy_density}
\end{figure}

Equation~\eqref{eq:superfluid_bjorken_energy_conservation} may be solved explicitly for \(\Th(g)\). The resulting expression for the local temperature \(T = \Th(g)/\t\) is
\begin{equation} \label{eq:gubser_superfluid_temperature_solution}
    T(\t,\xp) = \frac{3 n \tilde\S\le(g\ri)^{2/3} - k \xh^2}{6 n \t \tilde\S\le(g\ri)^{1/3}},
\end{equation}
where \(\tilde\Sigma\) is related to the function appearing in equation~\eqref{eq:ideal_bjorken_temperature} by
\begin{equation}
    \tilde\S(g) = \S\le(\frac{1}{2q\sqrt{1+g^2}}\ri) = \frac{1}{q^2\t_0^2 (1+g^2)} + \sqrt{\frac{1}{q^4 \t_0^4{(1+g^2)^2}} + \frac{ k^3 \xh^6}{27n^3}},
\end{equation}
where \(\t_0\) is an integration constant, chosen such that \(T\) reduces to the Bjorken flow result~\eqref{eq:ideal_bjorken_temperature} in the \(q \to 0\) limit. Substituting the solution for the local temperature into equation~\eqref{eq:model_energy_density} with \(\m=0\), we then obtain the energy density
\begin{equation} \label{eq:gubser_superfluid_energy_density_solution}
    \ve(\t,\xp) =  3 n \le[\frac{3n \tilde\S(g)^{2/3} -   k \xh^2}{6 n \t \tilde\S(g)^{1/3}}\ri]^4 + \frac{k \xh^2}{2 \t^2} \le[\frac{3 n \tilde\S(g)^{2/3} -   k \xh^2}{6 n \t \tilde\S(g)^{1/3}}\ri]^2 + \frac{f_\pi^2 \xh^2}{2 \t^2}.
\end{equation}
Since this is a rather complicated function of \(\t\) and \(\xp\), we plot the typical dependence of \(\ve\) on \(\xp\) of the energy density in figure~\ref{fig:example_gubser_energy_density}, for several values of \(\t\). The plots were made by setting~\(n = \pi^2/30\), \(k = 1/6\), \(q = f_\pi\), \(\xh = 1/10\), and \(\t_0 = 1/f_\pi\), the same values as in figure~\ref{fig:viscous_temperature}.

The solid blue curves in the figure show the result for the energy density~\eqref{eq:gubser_superfluid_energy_density_solution}, while the dashed orange curves show the same energy density, but with the \(\xp\)-independent contribution \(f_\pi^2 \xh^2/2\t^2\) subtracted. For comparison, we also show in dotted black the normal fluid result~\eqref{eq:gubser_normal_fluid_energy_density_solution}, with \(\t_0\) chosen such that the two energy densities agree at \(\xp=0\) for the earliest plotted time \(q\t =0.1\).

In both the superfluid and normal fluid cases, the energy density exhibits a single peak as a function of \(\xp\), which at early times \(q \t < 1\) is located at \(\xp = 0\). For \(q \t > 1\) the peak is located at \(q\xp = \sqrt{q^2 \t^2 - 1}\).\footnote{From the definition of \(g\) in equation~\eqref{eq:gubser_normal_fluid_energy_density_form}, we see that \(q \xp = \sqrt{q^2 \t^2 - 1}\) corresponds to \(g=0\).} Note that since \(\xp\) is the radial coordinate orthogonal to the beam axis, the peak at non-zero \(\xp\) is really a ring of energy density surrounding the beam axis, that expands outward with a speed asymptotically approaching the speed of light as \(\t \to \infty\). Away from the peak, the superfluid energy density is dominated by the \(\xp\)-independent contribution to~\eqref{eq:gubser_superfluid_energy_density_solution} proportional to \(f_\pi^2\).

The form of the late-time decay of the energy density depends on the value of \(\xp\). At the peak  at \(q\xp = \sqrt{q^2 \t^2 - 1}\), the energy density takes the relatively simple form
\begin{equation}
    \ve\le(\t,\sqrt{\t^2-q^{-2}}\ri) =  \frac{f_\pi^2 \xh^2}{2 \t^2} + \frac{C}{\t^4},
\end{equation}
where
\begin{equation}
    C = 3 n \le[ \frac{3 n \tilde{\S}(0)^{2/3} - k \xh^2}{6 n \tilde{\S}(0)^{1/3}} \ri]^4 + \frac{k \xh^2}{2 \t^2} \le[ \frac{3 n \tilde{\S}(0)^{2/3} - k \xh^2}{6 n \tilde{\S}(0)^{1/3}} \ri]^2.
\end{equation}
Away from this peak, for large \(\t\) at fixed \(\xp\) the energy density decays as\footnote{Note that the expansion~\eqref{eq:superfluid_gubser_late_time} diverges in the limit \(\xh \to 0\), despite the fact that the full expression for the energy density in equation~\eqref{eq:gubser_superfluid_energy_density_solution} is finite in this limit. The reason for this noncommutativity of limits is that the function \(\tilde{\S}(g)\) appearing in equation~\eqref{eq:gubser_superfluid_energy_density_solution} tends to zero in the \(\t \to \infty\) and \(\xh \to 0\) limit, at a speed that depends on the order of limits.}
\begin{equation} \label{eq:superfluid_gubser_late_time}
    \ve(\t,\xp) \approx \frac{f_\pi^2 \xh^2}{2\t^2} + \frac{8 n^2 }{k \xh^2  {\t_0}^4 q^8 \t^8} + \frac{32 n^2 (q^2 \xp^2-1)}{ k \xh^2 \t_0^4 q^{10}\t^{10}} + \cO(\t^{-12}).
\end{equation}
The dominant term in both cases is proportional to \(f_\pi^2/\t^2\) and is independent of \(\xp\), consistent with the behaviour plotted in figure~\ref{fig:example_gubser_energy_density}. We see that first subleading correction at late times decays more rapidly away from the peak.

\subsection{Dissipative corrections}
\label{sec:dissipative_gubser}

We now consider the effect of dissipative corrections on the superfluid Gubser flow of the previous subsection. The symmetry arguments fixing the velocities are unchanged from the ideal case, such that \(u\) is given by equation~\eqref{eq:gubser_normal_fluid_fluid_velocity}, and \(\x = \t^{-2} \xh \, \p_w\) with constant \(\xh\). With this form of \(\x^\m\) one finds \(\nabla_\m \le[f(T)^2 \x^\m\ri] = 0\), so that the transport ceofficient \(\z_3\) makes no contribution to the hydrodynamic equations for the flows that we are considering. We will consider only flows with \(\m = 0\), so the coefficient \(\k\) also makes no contribution.\footnote{If one were to consider a flow with \(\m \neq 0\), transforming with weight \(\a_\m=1\) as required by dimensional analysis, one would find \(\Delta^{\m\n} \nabla_\n \le(\frac{\m}{T}\ri) = 0\) for the fluid velocity in equation~\eqref{eq:gubser_normal_fluid_fluid_velocity}. As a result, \(\k\) again drops out of the hydrodynamic equations.}

As argued in section~\ref{sec:bjorken_dissipative}, of the three remaining transport coefficients \(\{\h,\z_1,\z_2\}\) we expect the shear viscosity \(\h\) to be most important for dissipation at small relative superfluid velocity. Moreover we expect that the shear viscosity will be dominated by the normal fluid contribution, such that it may be approximated by \(\h \approx \bar{\h} T^3\) for some dimensionless constant \(\bar{\h}\). Making this approximation, and neglecting the contributions of \(\z_{1,2}\) to the hydrodynamic equations, we find that the only independent hydrodynamic equation is
\begin{equation} \label{eq:superfluid_gubser_energy_conservation_dissipative}
	(1+g^2) \le[12 n \Th(g)^2 + k \xh^2\ri] \Th'(g) + 2 g \le[4 n \Th(g)^2 + k \xh^2 \ri]\Th(g) + \frac{4\bar{\eta}g^2}{3\sqrt{1+g^2}} T(g)^2= 0\,. 
\end{equation}

We are not able to solve equation~\eqref{eq:superfluid_gubser_energy_conservation_dissipative} analytically, however it is straightforward to solve numerically. To do so, we first solve the equation perturbatively for large values of $g$, finding
\begin{equation}\label{eq:superfluid_gubser_temperature_expansion}
	\hat T(g) = \frac{n}{q^2\tau_0^2 \xh^2 k} \frac{1}{g^2} + \frac{n \left(2 \bar{\eta} n- 3 k^2 \xh^4 q^2 \tau _0^2\right)}{3q^4
		\tau _0^4 k^3 \xh^6 } \frac{1}{g^4} + \dots\;,
\end{equation}
where $\tau_0$ is an integration constant, chosen such that the leading term in this expansion matches the leading term in the large-\(g\) expansion of the ideal solution~\eqref{eq:gubser_superfluid_temperature_solution}. We then use this expansion to set boundary conditions at some large value of \(g\), and then numerically integrate to smaller values of \(g\). The resulting solution for \(T = \Th(g)/\t\) can then be substituted into equation~\eqref{eq:model_energy_density} to obtain the energy density. 

\begin{figure}
    \definecolor{plotpurple}{rgb}{0.5,0,0.5}
    \begin{center}
        \hspace{45pt}\includegraphics{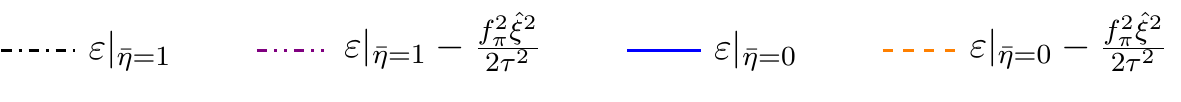}
        \vspace{-0.5cm}
        \includegraphics{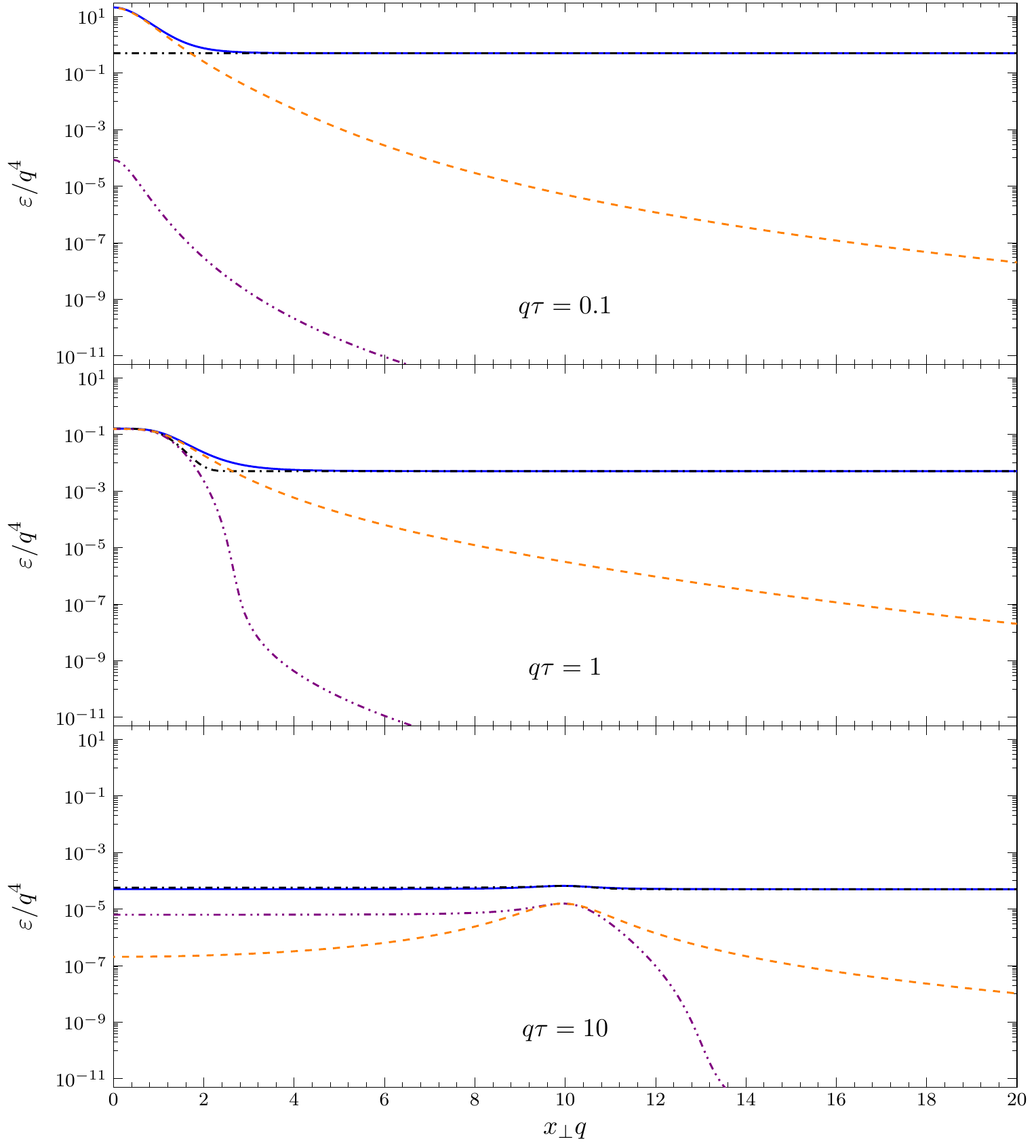}
    \end{center}
\caption{Sample logarithmic plots of the energy density of ideal and viscous superfluid Gubser flow as a function of transverse distance \(\xp\), at different Milne times \(\t\). The solid blue and dashed orange curves show the same data as in figure~\ref{fig:example_gubser_energy_density}. The dot-dashed black curve shows the numerical result for the energy density at \(\bar{\h} = 1\), with \(\t_0 \approx 74/f_\pi\) and otherwise the same values of parameters as the ideal fluid results. The dot-dot-dashed purple curve shows the \(\bar{\h}=1\) result, but with the \(\xp\)-independent contribution \(f_\pi^2 \xh^2/2\t^2\) subtracted, in order to make the \(\xp\)-dependence clearer.}
\label{fig:viscous_energy}
\end{figure}

For example, in figure~\ref{fig:viscous_energy} we show the result of this numerical procedure for a sample value of the dimensionless shear viscosity $\bar\eta = 1$, together with the inviscid solutions shown in figure~\ref{fig:example_gubser_energy_density}. The rest of the parameters in the model were set to the same values as in the latter figure, namely \(n = \pi^2/30\), \(k = 1/6\), \(q = f_\pi\), \(\xh = 1/10\). For the \(\bar{\h}=0\) curves we used \(\t_0 = 1/f_\pi\), while for the \(\bar{\h}=1\) curves we tuned \(\t_0 \approx 74/f_\pi\), such that the value of the energy density at the local maximum (as a function of \(\xp\)) matches the \(\bar{\h}=0\) result at late times. From the figure, we see that non-zero shear viscosity tends to lead to reduced energy density at early times, and much more rapid decay of the energy density at large \(\xp\). These two effects are related by the conformal symmetry: from the form of \(g\) in equation~\eqref{eq:gubser_normal_fluid_energy_density_form} we see that both small \(\t\) and large \(\xp\) correspond to large positive \(g\).

As for Bjorken flow, we find that the superfluid component regularises the pathology pointed out in ref.~\cite{Gubser:2010ze}. When $\xh =0 $, equation~\eqref{eq:superfluid_gubser_energy_conservation_dissipative} may be solved to obtain~\cite{Gubser:2010ze}
\begin{equation}\label{eq:gubser_temperature}
	\Th(g)  =  \frac{q\ve_0^{1/4}}{\left(1+g^2\right)^ {1/3}} + \frac{ \bar{\eta} g }{3n\left(1+g^2\right)^ {1/2}}\left[1 -  \left(1+g^2\right)^ {1/6}  {}_2F_1\left(\frac{1}{2},\frac{1}{6};\frac{3}{2};-g^2\right)\right],
\end{equation}
with integration constant \(\ve_0\), leading to the energy density~\eqref{eq:energy_Gubser_viscous}. This solution for \(\Th(g)\) becomes negative at large positive \(g\), corresponding to small \(\t\) or large \(\xp\), signalling a breakdown of hydrodynamics in these regimes since the viscous contribution outweighs the ideal fluid result. No such behaviour is visible in the superfluid energy density plotted in figure~\ref{fig:viscous_energy}, where a change in the sign of \(\Th\) would lead to a zero in the subtracted energy density (the dot-dot-dashed purple curve in the figure).

\section{Comments on introducing a pion mass}
\label{sec:mass}

In QCD, chiral symmetry is only approximate, giving a small mass \(m\) to the pion \(\vf\). To what extent do the massless flows found in sections~\ref{sec:bjorken} and~\ref{sec:gubser} provide good approximations to the true flows for small non-zero \(m\)?

A necessary condition for the validity of the approximation is that the kinetic term in the pion Lagrangian dominates the mass term, \((\p \vf)^2 \gg m^2 \vf^2\), when evaluated on the \(m=0\) solution. All of our massless solutions have \(\vf = \xh w\), and so this condition is equivalent to \(m^2 w^2 \t^2 \ll 1\), where the factor of \(\t^2\) arises from \(g^{ww} = \t^{-2}\) in the metric~\eqref{eq:metric}. For given values of \(m\) and \(w\), this implies an upper bound on \(\t\), beyond which we cannot trust the massless-pion approximation.

This is dangerous, since we would typically only trust hydrodynamics at late times where we expect gradients to become small. For instance, when the pion is the real pion of QCD with \(m \approx 1.4 \,\mathrm{fm}^{-1}\) in natural units, we expect the massless-pion approximation to hold for
\begin{equation}
    \t^2 \ll \frac{(1 \, \mathrm{fm/c})^2}{w^2},
\end{equation}
where we have restored a factor of the speed of light \(c\). Since \(1 \, \mathrm{fm}/c\) is of the order of the time scale for hydrodynamisation in heavy-ion collisions~\cite{Schlichting:2019abc}, one cannot hope to accurately describe the regions with \(|w| \sim \cO(1) \) by setting \(m=0\). Instead, we expect that the massless pion approximation will only be accurate for sufficiently small values of the rapidity \(w\).

We will now estimate the size of massive corrections to superfluid Bjorken flow in more detail. Throughout this section we consider only ideal hydroydnamics, for simplicity. A small pion mass modifies the conservation equation of the \(U(1)\) current to
\begin{equation} 
    \label{eq:conservation_law_mass}
    \nabla_\m J^\m = f^2 m^2 \sin \vf,
\end{equation}
while the Josephson condition and conservation of \(T^{\m\n}\) are unchanged. Using the constitutive relations written in equation~\eqref{eq:ideal_hydro_tensors} with \(\r = \chi \m\) and \(\x^\l = \p^\l \vf - \m u^\l\) the conservation equation becomes
\begin{equation} \label{eq:conservation_law_mass_explicit}
    \nabla_\l \le[f^2 \p^\l \vf +  (f^2 - \chi) u^\l u^\a \p_\a \vf\ri] = f^2m^2 \sin \vf,
\end{equation}
where we have used the Josephson condition to replace \(\m\) with \(-u^\a \p_\a \vf\).

The dependence of \(f^2\) and \(\chi\) on the local temperature couples this equation to the conservation equation for the stress tensor. We have not been able to find explicit solutions to the resulting set of coupled equations for the hydrodynamic variables, so for simplicity we will approximate the susceptibilities by their zero temperature value \(f^2 \approx \chi \approx f_\pi^2\). The conservation law~\eqref{eq:conservation_law_mass_explicit} then becomes simply the sine-Gordon equation
\begin{equation} \label{eq:sine_gordon}
    \Box \vf= m^2 \sin \vf
\end{equation}
This approximation is not very interesting from the point of view of hydrodynamics, since we have essentially decoupled the pions from the hydrodynamic flow. However, it is hopefully good enough to estimate the size of massive corrections to our solutions.

We will solve equation~\eqref{eq:sine_gordon} perturbatively at small \(m\) by expanding the pion field as 
\begin{equation} \label{eq:pion_massive_expansion}
    \vf(\t,w) = \xh w + m^2 \vf_1(\t,w) + \cO(m^4),
\end{equation}
where \(\xh w\) is the \(m=0\) solution found in sections~\ref{sec:bjorken} and~\ref{sec:gubser}, and \(\vf_1\) is the leading-order correction at non-zero \(m\). Substituting this expansion into equation~\eqref{eq:sine_gordon} and discarding terms \(\cO(m^4)\) and higher, one finds
\begin{equation}
    - \t \p_\t \le(\t \p_\t \vf_1 \ri)  + \p_w^2 \vf_1 = \t^2 \sin (\xh w).
\end{equation}
The general solution to this sourced wave equation is
\begin{equation} \label{eq:massive_pion_solution_general}
    \vf_1(\t,w) = W_+(\t e^w) + W_-(\t e^{-w}) - \frac{\t^2}{4+\xh^2} \sin(\xh w),
\end{equation}
where \(W_\pm(\t^{\pm w})\) are left- and right-moving waves along the beam axis.\footnote{Moving to the cylindrical coordinate system~\eqref{eq:metric_cylindrical} by setting \(\t = \sqrt{t^2 - z^2}\) and \(w = \frac{1}{2} \log\le(\frac{t+z}{t-z}\ri)\), one finds \(\t e^{\pm w} = t \pm z\).} Note that the non-linear dependence of \(\vf_1\) on \(w\) indicates that boost-invariance is broken by non-zero~\(m^2\).

The relative size of the \(\cO(m^0)\) and \(\cO(m^2)\) contributions to \(\vf\) depends on the explicit form of the functions \(W_\pm\). We will motivate a particular form of these functions by demanding that the contribution of the pion to the stress tensor is approximately boost-invariant near \(w=0\). Concretely, substituting the expansion~\eqref{eq:pion_massive_expansion} into the sine-Gordon stress tensor \(T^{\m\n} = f_\pi^2 \le[\p^\m \vf \p^\n \vf - \frac{1}{2} g^{\m\n} (\p\vf)^2 + m^2 g^{\m\n} \cos \vf \ri]\), we find the non-zero components
\begin{align}
    T^{\t\t} &= \frac{f_\pi^2 \xh^2}{2\t^2} + f_\pi^2 m^2 \le[ \frac{\xh \p_w \vf_1}{\t^2} - \cos(\xh w) \ri] + \cO(m^4),
    \nonumber \\
    T^{\t w} &= - \frac{f_\pi^2 m^2\xh \p_\t \vf_1}{\t^2} + \cO(m^4),
    \\
    T^{ww} &= \frac{f_\pi^2 \xh^2}{2\t^4} + \frac{f_\pi^2}{\t^2} \le[\frac{\xh \p_w\vf_1}{\t^2} + \cos(\xh w) \ri] + \cO(m^4),
    \nonumber
\end{align}
and \(  T^{\xp\xp} =  \xp^2 T^{\q\q} = - T^{\t\t}\). If we impose approximate boost invariance near \(w=0\) by setting \(\p_w T^{\m\n}|_{w=0} = 0\), this yields the conditions \(\p_\t \p_w \vf_1|_{w=0} = \p_w^2 \vf_1|_{w=0} = 0\). Substituting the solution~\eqref{eq:massive_pion_solution_general} into these conditions yields two second-order ordinary differential equations for \(W_\pm\), which may be solved to obtain
\begin{equation} \label{eq:massive_pion_solution_general}
    \vf_1(\t,w) = c \log \t + \frac{\t^2}{4+\xh^2} \le[\sin(\xh w) - \frac{\xh}{2} \sinh(2 w) \ri],
\end{equation}
where \(c\) is an integration constant.\footnote{There is only one integration constant in equation~\eqref{eq:massive_pion_solution_general} because we have set boundary conditions to eliminate a term which is independent of \(\t\) and linear in \(w\) --- since such a term can be trivially removed by a redefinition of the \(\cO(m^0)\) integration constant \(\xh\) --- and a constant term independent of both \(\t\) and \(w\) that may be removed by the shift symmetry present when \(m=0\).}

If we further impose the physically reasonable condition that \(\vf \to 0\) as \(t \to \infty\) for fixed \(z\), then we find \(c = 0\) in equation~\eqref{eq:massive_pion_solution_general}. Then, expanding \(\vf_1(\t,w)\) for small \(w\) one finds \(m^2\vf_1(\t,w) \approx - \xh m^2 \t^2 w^3/6\). The massless-pion approximation will be self-consistent only if this is much smaller in magnitude than the \(m=0\) solution \(\vf_0 = \xh w\), and so we recover the estimate \(m^2 \t^2 w^2 \ll 1\) made at the beginning of this section. We emphasise that this conclusion depends strongly on the functions \(W_\pm\), however. For instance, if one were to choose boundary conditions that set \(W_\pm = 0\), and also assume that \(\xh \ll 1\), then to leading order at small \(\xh\) one finds instead \( |m^2 \vf_1/\vf_0| \sim m^2 \t^2\).

\section{Discussion}
\label{sec:discussion}

In this work we have presented a number of boost-invariant solutions to the equations of motion of superfluid hydrodynamics, for a superfluid with a simple, yet physically motivated equation of state~\eqref{eq:model_equation_of_state}. In the superfluid generalisation of Bjorken flow presented in section~\ref{sec:bjorken}, we found that the superfluid component of the flow leads to subleading terms in the energy density at late times beginning at \(\cO(\t^{-2})\), the same order in the late-time expansion as viscous effects appear.

In section~\ref{sec:gubser}, by applying Gubser's conformal symmetry based approach~\cite{Gubser:2010ze}, we found flows expanding transverse to the boost direction. The superfluid component of the flow provides a contribution to the energy density that is independent of the transverse direction and dominates at late times, plus additional subleading corrections to the flow at large transverse direction. The former contribution is rather physically unsatisfying, and in future work it would be pleasing to find a modification of the flow in which the superfluid contribution to the energy density is also expanding.

To find solutions to the hydrodynamic equations, we have treated the spontaneously broken symmetry giving rise to the superfluid as exact, so that the resulting Goldstone bosons are massless. In QCD chiral symmetry is only approximate, and thus the pions have non-zero masses, with important consequences for transport. In section~\ref{sec:mass} we estimated the size of massive corrections to the above flows, concluding that the massless-pion approximation should hold for the region of the flow at small spatial rapidities \(w \ll 1\). A more complete treatment of our flows would include the pion masses throughout. However, one would need to take into account the complication that a non-zero pion mass is incompatible with boost invariance.

There are many other possible directions for further work. For instance, one could study second-order dissipative effects, or perturbatively study anisotropy in the transverse plane (as would be introduced by non-zero impact parameter in a heavy-ion collision) using the methods of ref.~\cite{Gubser:2010ui}.

Throughout this work we have considered only superfluids arising from the spontaneous breaking of a \(\mathrm{U}(1)\) symmetry. As discussed in section~\ref{sec:intro}, our solutions apply also to non-abelian superfluids, with our \(\mathrm{U}(1)\) being a subgroup of the full non-abelian symmetry.\footnote{Here we assume that the pion susceptibility \(f^2\) is diagonal in the broken symmetry group indices. A case with non-diagonal \(f^2\) was studied in ref.~\cite{Hoyos:2014nua}.} A natural question is whether a non-abelian superfluid permits more general boost-invariant flows with multiple pions excited.

One might attempt to generalise superfluid Bjorken flow along the lines of ref.~\cite{Gubser:2012gy}, in which a complex time translation was applied to the Bjorken stress tensor, resulting in a complexified stress tensor \(T^{\m\n}_\mathbb{C}\). Since the hydrodynamic equations are invariant under time translations and linear in the conserved currents, \(T^{\m\n} \equiv \Re T^{\m\n}_\mathbb{C}\) remains an exact solution. For an appropriately chosen translation, the new stress tensor is approximately that of Bjorken at small rapidities, while providing a better fit to data at large rapidities. Applying the aforementioned procedure to the stress tensor of superfluid Bjorken flow would lead to a conserved \(T^{\m\n}\), however, it seems unlikely that this would be the stress tensor of a superfluid: since the superfluid stress tensor in equation~\eqref{eq:ideal_hydro_tensors} is non-linear in \(\x^\m\), the relative superfluid velocity that would be read off from this new \(T^{\m\n}\) would probably not be expressible as \(\x^\m = \Delta^{\m\n} \p_\n \vf\) for some scalar \(\vf\).

Finally, as mentioned in section~\ref{sec:intro}, possibly the largest barrier to  the physical interpretation of our results is the restriction to temperatures much smaller than \(f_\pi\), clashing with the chemical and kinetic freezeouts that occur at low temperatures in heavy-ion collisions. It is thus desirable to extend our treatment to larger temperatures by adding new terms to the equation of state~\eqref{eq:model_equation_of_state}. Such terms would further break conformal symmetry, so we do not expect it to be possible to repeat the analysis of section~\ref{sec:gubser}. It should be straightforward to extend the translationally-invariant case discussed in section~\ref{sec:bjorken}, although solving the resulting equations may require numerics. An additional complication is that sufficiently close to \(T_c\) one must also take into account fluctuations of the order parameter, taking one to the regime studied in ref.~\cite{Mitra:2020hbj}.

We hope that the solutions presented in this work, and perhaps some of the generalisations that we have suggested, provide helpful intuition for superfluid effects in highly relativistic hydrodynamic flows, and that they may be a useful tool for benchmarking future numerical codes.

\acknowledgments

We would like to thank Carlos Hoyos, Alexander Krikun, Niels Obers, and Konstantin Zarembo for useful discussions and comments on a draft of this manuscript. We would particularly like to thank Konstantin Zarembo for highlighting the open problem of superfluid Gubser flow. Nordita is supported in part by NordForsk.

\appendix

\bibliographystyle{JHEP}
\bibliography{superfluid_ref}

\end{document}